\documentclass[twocolumn,superscriptaddress,nofootinbib,preprintnumbers,showpacs,tightenlines,notitlepage]{revtex4-1}
\usepackage{graphics}
\usepackage{amsmath}
\usepackage{color}

\usepackage{graphicx}
\usepackage{amssymb}
\usepackage{float}
\usepackage{amsmath}
\usepackage{amsfonts}
\usepackage{dcolumn}
\usepackage{amsthm}
\usepackage{bm}

\def\3nab{\tilde{\nabla}}

\def\be {\begin{equation}}
\def\ee {\end{equation}}
\def\ba {\begin{eqnarray}}
\def\ea {\end{eqnarray}}

\newcommand{\sfr}[2]{{\textstyle\frac{#1}{#2}}}

\newcommand{\barray}{\begin{array}}
\newcommand{\earray}{\end{array}}
\newcommand{\e}{e}

\let\a=\alpha      \let\e=\varepsilon
     \let\th=\theta  \let\l=\lambda
\let\m=\mu             \let\p=\pi    
 \let\t=\tau   \let\f=\phi \let\ph=\varphi
   \let\o=\omega
\let\G=\Gamma \let\D=\Delta   
         
\let\O=\Omega 
\let\W=\wedge

\def\\{\hfill\break} \let\==\equiv

\let\0=\noindent

\def\qed{\hfill\raise1pt\hbox{\vrule height5pt width5pt depth0pt}}

\def\be{\begin{equation}}
\def\ee{\end{equation}}
\def\bea{\begin{eqnarray}}\def\eea{\end{eqnarray}}
\def\lt({\left(} \def\rt){\right)}

\begin{document}

\title{Direct Detection of Universal Expansion by Holonomy in the McVittie Spacetime}

\author{Tony Rothman}
\email{rothman@nyu.edu}
\affiliation{Department of Applied Physics, New York University}
\author{Mariam Campbell}
\email{CMPMAR009@myuct.ac.za}
\affiliation{Department of Mathematics and Applied Mathematics and ACGC, University of Cape Town,
Cape Town, Western Cape, South Africa.}
\author{Rituparno Goswami}
\email{Goswami@ukzn.ac.za}
\affiliation{Astrophysics and Cosmology Research Unit, School of Mathematics, Statistics and Computer Science, University of KwaZulu-Natal, Private Bag X54001, Durban 4000, South Africa.}
\author{George F.R. Ellis}
\email{George.Eliis@uct.ac.za}
\affiliation{Department of Mathematics and Applied Mathematics and ACGC, University of Cape Town,
Cape Town, Western Cape, South Africa.}

\begin{abstract}
In general relativity the parallel transfer of a vector around a closed curve in spacetime, or along two curves which together form a closed loop, usually results in a nonzero deficit angle between the vector's initial and final positions. We show that such holonomy in the McVittie spacetime, which represents a gravitating object imbedded in an expanding universe, can in principle be used to directly detect the expansion of the universe, for example by measuring changes in the components of a gyroscopic spin axis.  Although such changes are of course small, they are large enough  ($\D S \sim 10^{-7}$) that they could conceivably be measured if the real universe behaved like the McVittie spacetime.  The real problem is that virialization will lead to  domains decoupled from the global expansion on a scale much larger than that of the solar system, making such an experiment infeasible probably even in principle. Nevertheless the effect is of interest in relation to ongoing discussions, dating back at least to Einstein and Straus, which concern the relationship between the expansion of the universe and local systems. \\
 \end{abstract}

\maketitle

\section{Introduction}
\setcounter{equation}{0}\label{sec1}

Attempts to understand how the large-scale behavior of the universe might affect local physics have been ongoing at least since Einstein attempted to incorporate Mach's Principle into general relativity. Einstein himself continued this line of inquiry in his paper with Straus on modeling a Schwarzschild domain in an expanding universe\cite{ES45}, and this work was itself further generalized in various ways by other authors (see e.g.\cite{NP71}). Recent papers on the same general theme have included Bochiccio and Faraoni\cite{BF12}, who examine how a Friedmann-Lema\^itre-Roberston-Walker (FLRW) cosmology affects the behavior of a Lema\^itre-Tolman-Bondi system; Faraoni and Jacques\cite{FJ07}, who examine whether whether various systems embedded in a FLRW cosmology participate in the expansion; and Cooperstock, Faraoni and Vollick\cite{CFV98}, who ask how the universal expansion of an FLRW universe affects the equations of motion in a local inertial frame.

One of the earliest and most important investigations in this area was, of course, that of McVittie\cite{McV33}, who discovered a solution to the Einstein equations that represents a spherically symmetric object in an expanding universe.  For the past two decades there has been some renewed interest in McVittie's solution after it was realized that many misstatements about the metric have been made in the literature\cite{Nolan98-99} and that a proper understanding of the spacetime was much more subtle than previously thought\cite{KKM10,LA10}.

The various controversies involving the horizon structure and nature of the central object in the McVittie solution do not concern us in the present investigation, which more closely resembles \cite{BF12}-\cite{CFV98}.  We merely intend to use the McVittie spacetime as a background to ``design" a few simple thought experiments that could, in principle, directly detect the expansion of the universe through the holonomy produced by the metric.  That is,  parallel transport of a vector around a closed loop in a curved spacetime generally results in a measurable deficit angle between the initial and final directions of the vector.  Rothman, Ellis and Murugan\cite{REM01} (REM) calculated the deficit angle produced for a variety of trajectories in the Schwarzschild-Droste\footnote{Johannes Droste, a pupil of Lorentz, independently announced the Schwarzschild exterior solution
within four months of Schwarzschild\cite{Droste}.}
static geometry and showed that this metric produces a quantized band structure of holonomy invariance. These results were generalized by Maartens, Mashhoon and Matravers to stationary axisymmetric spacetimes\cite{MMM02}.  In the current paper we carry out an analysis similar to REM's for the McVittie metric.  Any cosmological expansion should affect the deficit angle of a vector under parallel transport, in principle allowing direct experimental detection of the universe's expansion. Of course, one expects such effects to be extremely small, and they are, but they turn out to be surprisingly large compared to, for example, the dimensionless strain of $10^{-21}$ successfully measured by LIGO.

 This paper is organised as follows: In the next section we discuss the basic geometry of McVittie spacetime and write down the parallel transport equations in an orthonormal tetrad basis. In Section 3, we discuss the holonomy for the vectors parallel transported along circular orbits in equatorial plane. Circular orbits are not actually geodesics in the McVittie cosmology, but as we show the error introduced by using such orbits as proxies for geodesics in computing the holonomy is negligible. As explained in detail, such an experiment require two measuring devices (gyroscopes) to be sent along different paths to meet at the same spatial location where their spin-axis directions can be compared, thus directly measuring holonomy. Two different versions are considered: an experiment with  one comoving and one orbiting apparatus (Section \ref{sec3:dual}); and an experiment with two counter orbiting gyroscopes (Section \ref{sec:3_counter}). In Section 4 we consider the outcome, which depends on the scale at which static or quasi-static domains coalesce out of the expanding universe as structure formation takes place.

\section{McVittie Metric}
\label{sec2}

The line element for the McVittie spacetime in isotropic coordinates is given by Faraoni\cite{Faraoni}, Eq. (4.10):
\bea\label{mcv}
ds^2 &=& -\frac{\left[1-\frac{m_o}{2a(t)r}\right]^2}
 {\left[1+\frac{m_o}{2a(t)r}\right]^2} dt^2\nonumber\\
 &&+a^2(t)\left[1 + \frac{m_o}{2a(t)r}\right]^4(dr^2 + r^2d\O^2)\, ,
\eea
where $r$ here corresponds to $\tilde r$ in Faraoni, $a(t)$ is the cosmological scale factor, and $m_o$ the mass of the central object. Note that $a  = 1$ gives the Schwarzschild solution and $m_o = 0$ gives the flat FLRW universe.  Thus, the general interpretation that the metric represents a central object in an expanding universe.

We have chosen this form of the metric (as opposed to the more common ``canonical nondiagonal form" in terms of an areal radius; see Faraoni Eq. (4.16)) for the important reason that we wish to use a radial coordinate that is tied to the matter.  If we send out, for example, a 
gyroscope along a closed loop, we want to measure the deficit angle in the components by the same material ``apparatus".  In Eq. (\ref{mcv}) $r$ represents a comoving coordinate; that is, a given matter particle (galaxy) is attached to a given $r$ because the normalised 4-velocity \begin{equation}\label{eq:4_vel}
u^a = \frac{[1+\frac{m_0}{2a(t)r}]}{[1-\frac{m_0}{2a(t)r}]} \delta^a_0
\end{equation} is a Ricci eigenvector, and hence an eigenvector of the matter stress tensor $T_{ab}$.
This represents the average motion of matter at each spacetime event, and hence corresponds to the idea of a fundamental observer in cosmology.
Thus, a gyroscope traveling along a circular orbit will return to the original apparatus if $r = constant$ but not if the proper distance $d(t) = ra(t)= constant$; in that case the universe has expanded during the transit time and the gyroscope will return to a different device. Hence in this paper, ``circular'' means circular in comoving coordinates (\ref{mcv})-(\ref{eq:4_vel}).\\

We perform all our calculations in a orthonormal tetrad basis $\{e_a \}$, which correctly describes the local physics.  The obvious choice for such a tetrad for metric (\ref{mcv}), as represented by the dual basis 1-forms $\{\omega^a\}$, is
\bea
\o^0=\o^t &=& \frac{\left[1-\frac{m_o}{2a(t)r}\right]}
 {\left[1+\frac{m_o}{2a(t)r}\right]} dt\label{omega1}\\
\o^1 =\o^r &=&  a(t)\left[1 + \frac{m_o}{2a(t)r}\right]^2 dr\label{omega2}\\
\o^2 = \o^\th &=&  a(t)\left[1 + \frac{m_o}{2a(t)r}\right]^2 r d\th \\
\o^3= \o^\ph & =& a(t)\left[1 + \frac{m_o}{2a(t)r}\right]^2 r\sin\th d\ph.
\eea
Working out the connection coefficients by the Cartan equation $d\o^a = -\o^a_{\,b} \W\o^b$ gives
\bea
\o^0_{\ 1}&=&\o^1_{\ 0} = \frac{m_o}{r^2 a^2\lt(1+\frac{m_o}{2ra}\rt)^2\lt(1-\frac{m_o^2}{4r^2a^2}\rt)}\, \o^0\nonumber\\
&&+\frac{\dot a}{a} \, \o^1\label{omega01}\\
\o^0_{\ 2} &=& \o^2_{\ 0}=\frac{\dot a}{a}\, \o^2 \label{omega02}\\
\o^0_{\ 3} &=& \o^3_{\ 0}=\frac{\dot a}{a} \, \o^3 \label{omega03}\\
\o^2_{\ 1} &=& -\o^1_{\ 2}=  \frac{\left(1 - \frac{m_o}{2ra}\rt)} {ar\lt(1+ \frac{m_o}{2ra}\rt)^3} \, \o^2\label{omega21} \\
\o^3_{\ 1} &=& -\o^1_{\ 3}=  \frac{\left(1 - \frac{m_o}{2ra}\rt)} {ar\lt(1+ \frac{m_o}{2ra}\rt)^3} \, \o^3\label{omega21} \\
\o^2_{\ 3} &=& -\o^3_{\ 2}= \frac{\cot\th}{\lt(1+ \frac{m_o}{2ra}\rt)^3} \, \o^3\label{omega23}
\eea

The parallel transport equation is
\be
dA^{a} + \o^{a}_{\ {b}} A^{b} = 0, \label{PT}
\ee
which gives the change in a vector with tetrad components $A^a$ along a curve $x^b(\l)$ with  tangent vector $X^b(\l) = dx^b/d\l$ and curve parameter $\l$.  For $\l= t$ and $H \equiv \dot a/a$, Eq. (\ref{PT}) yields the following ordinary differential equations:
\begin{widetext}
\bea
dA^t &+& \frac{m_o}{a^2r^2\lt(1+ \frac{m_o}{2ra}\rt)^4}A^r dt
+ Ha\lt(1+\frac{m_o}{2ra}\rt)^2 A^r dr\nonumber\\
&+&Har\lt(1+\frac{m_o}{2ra}\rt)^2 A^\th d\th
+ Har\lt(1+\frac{m_o}{2ra}\rt)^2 \sin\th A^\f d\f=0 \label{dAt}\\
dA^r &+& \frac{m_o}{a^2r^2\lt(1+ \frac{m_o}{2ra}\rt)^4}A^t dt
+ Ha\lt(1+\frac{m_o}{2ra}\rt)^2 A^t dr\nonumber\\
&-& \frac{\lt(1-\frac{m_o}{2ra}\rt)}{\lt(1+\frac{m_o}{2ra}\rt)} A^\th d\th
  -\frac{\lt(1-\frac{m_o}{2ra}\rt)}{\lt(1+\frac{m_o}{2ra}\rt)}\sin\th A^\f d\f=0 \label{dAr}\\
dA^\th &+& H ar \lt(1+\frac{m_o}{2ra}\rt)^2 A^t d\th+ \frac{\lt(1-\frac{m_o}{2ra}\rt)}{\lt(1+\frac{m_o}{2ra}\rt)} A^r d\th
+\frac{cos\th\, (ar)}{\lt(1+\frac{m_o}{2ra}\rt)}A^\f d\f = 0\label{dAth}\\
dA^\f &+& H ar \lt(1+\frac{m_o}{2ra}\rt)^2 \sin\th A^t  \, d\f + \frac{\lt(1-\frac{m_o}{2ra}\rt)}{\lt(1+\frac{m_o}{2ra}\rt)} \sin\th A^r d\f
-\frac{cos\th\, (ra)}{\lt(1+\frac{m_o}{2ra}\rt)}A^\th d\f = 0.\label{dAf}
\eea
\end{widetext}
Note that $t$ is not an affine parameter, but that does not matter for our purposes; see the next section. {Indeed the curve with tangent vector $X^b$ need not even be a geodesic.}\\

We note that for parallel transport along any curve, magnitude is conserved:
\be\label{eq:const}
A^a g_{ab}A^b =  (A^a g_{ab}A^b)_o = const,
\ee
and so in the tetrad basis
\bea\label{constraint1}
-(A^t)^2 + (A^r)^2 + (A^\theta)^2 + (A^\phi)^2 = const.
\eea
We will make extensive use of this property in the following sections.

\section{Circular Holonomy}\label{sec3}
\subsection{Circular Orbits and Kepler's Law}\label{ss3.1}

We consider the  holonomy of vectors moving on a circular orbit in the equatorial plane.  It is important to point out that due both to a nonzero pressure gradient in the McVittie spacetime and the expansion of spacetime, circular orbits are not actually geodesics.  Rather, in our coordinates, over an orbital period particles spiral inward with time from a radius $r_1$ to $r_2$.  Because the geodesics are not closed, one cannot in principle measure holonomy on them unless  a force is exerted to ensure that any apparatus is somehow returned to its initial spatial location (which will be a very small displacement, as we show below). Likewise, an instrument will not follow a circular orbit without employing rockets to hold it at a fixed radius.  The use of rockets would, of course, introduce positioning errors into any experiment, which might very well overwhelm the desired results.

On the other hand, over an orbital period one can allow the apparatus to freely follow the geodesic from $r_1$ to $r_2$.  In the McVittie spacetime the difference between $r_1$ and $r_2$ is so small, however, that one introduces a negligible error by computing the holonomy as if the instrument were following a circular orbit.  We take this approach and show in \S \ref{sec:3_error} below that the error in measuring the holonomy is indeed negligible.

 {Below we will consider an experiment involving gyroscopes.  One might object that spinning objects do not follow geodesics, due to the coupling of the spin tensor to the Riemann curvature tensor, as manifested in the Mathisson-Papapetrou-Dixon equations.  This effect is entirely negligible for spinning bodies of less than astrophysical size.  Cornadesie and Papapetrou\cite{CP51} show that the ratio of the spin terms to the ordinary relativistic terms, which we consider, is $\sim (R^2/r^2)(T/\t)$, where $R$ is the size of the object, $r$ is the Schwarzschild coordinate, $T$ is the rotation period of the object around the Sun and $\t$ is the rotation period around its axis.  For a gyro of $R = 1$ m and $\t = 10^{-3}$ s at $r =$ 1 AU, this ratio is $\sim 10^{-12}$, utterly negligible.  In other words, for any gyroscope that can be treated as a point particle, the coupling vanishes.}

We further note that a  geodesic is generated whenever   the tangent vector to the curve remains parallel to itself.  Then the geodesic equation for an arbitrary curve parameter $\lambda$ is given by
\be\label{geoeqn}
D_{\partial/\partial\lambda}X^a= fX^a\Rightarrow X^b\nabla_bX^a=f X^a\;.
\ee
In terms of an  affine parameter of the geodesic, one will have $f=0$. For  the case when the proper time $\tau$ along the curve is the curve parameter, the function $f$ is
\be
f=\frac{d^2\lambda/d\tau^2}{(d\lambda/d\tau)^2}\;=0;
\ee
as is well known, proper time is an affine parameter. However if we choose the coordinate time $t$ to be the curve parameter, then for radius $r=constant$, $t$ is related to the proper time $\tau$ by
\be
d\tau=\frac{\left(1-\frac{k}{2a}\right)}{\left(1+\frac{k}{2a}\right)}dt,
\ee
where
\be
k\equiv \frac{m_o}{r}\;.
\ee
Calculating  $f$ we find
\be\label{eqf}
f=-\frac {{Hk}/{a}}{\left(1-\frac{k}{2a}\right)\left(1+\frac{k}{2a}\right)}.
\ee
Since $H = \dot a/a$, this equation shows that $f=0$ for $a=1$ or for $k=0$. The first situation corresponds to the Schwarzschild spacetime while the later corresponds to FLRW spacetime. Thus, for both the limiting cases of McVittie spacetime, the time coordinate $t$ is an affine parameter. However, it is not an affine parameter for the general case when both $k$ and $H$ are non-zero. Nevertheless, we can still use the coordinate time $t$ as the curve parameter along circular orbits because the parallel transport equation of vector $A^a$ moving  in the equatorial plane will not change:
\be\label{para}
X^b\nabla_b A^a=0\;.
\ee
Furthermore, taking $t$ as the curve parameter will enable us to compute the limiting cases easily, without performing complicated coordinate transformations.

For circular orbits as we have defined them, the comoving radial coordinate $r = constant$ (not the proper distance $d=ra$), which implies that $dr=0$.  The radial component of the tangent vector will be zero, so
\be \{r = r_0\}\Rightarrow\{X^r =0 \}\Rightarrow \{dX^r =0\}.\ee  Further, by symmetry we may take $\{\th =\pi/2\}$ and so
\be
  \{d\th =0\} \Rightarrow \{X^\th =0\} \Rightarrow \{dX^\th =0\}.
\ee
Therefore, with $t$ as the curve parameter, the components of the tangent vector in the tetrad frame are
\be
X^{\m} = \frac{1}{\alpha} \left[\frac{\lt(1-\frac{k}{2a}\rt)}
{ \lt(1+\frac{k}{2a}\rt)}, 0, 0,
ar\lt(1 + \frac{k}{2a}\rt)^2 \O\right]\label{tanvec}
\ee
where the normalization factor $\a$ is found by setting $X_\mu X^\mu = -1$, and the angular velocity is
\be
\O \equiv \frac{d\f}{dt}.
\ee
When $A^a$ is a tangent vector $X^a$, Eq. (\ref{dAr}) directly gives an algebraic relationship between $X^t$ and $X^\f$ :
\be
X^\f = \frac{k}{ra^2}\lt(1-\frac{k}{2a}\rt)^{-1}\lt(1+\frac{k}{2a}\rt)^{-3} \O^{-1}\, X^t .\label{AtAf}
\ee
Then Eqs. (\ref{tanvec}) and (\ref{AtAf})  immediately yield
\be
\O^2=\frac{k}{a^3r^2\lt(1+\frac{k}{2a}\rt)^6} \label{kepler}
\ee
independent of normalization, which is Kepler's Third Law for the McVittie spacetime in these coordinates, except that as already mentioned these are not strictly geodesic orbits. Over cosmological times $\O$ and hence the angular momentum of a body on a circular orbit would change.

We see that for $a=1$ (Schwarzschild), the orbital frequency $\O$ differs from the ``Newtonian" value  $\O_N \equiv \sqrt{k}/r$  by $\sim k \sim 10^{-8}$ in the vicinity of Earth.  In principle one could measure this deviation, assuming one could correct for other major perturbations. The additional effect due to the expansion of the universe would be even smaller.  For example, for a deSitter universe with $a = e^{Ht}$, $Ht << 1$ and $k << 1$, one has from Eq. (\ref{kepler}) to first order in small quantities
\be
\frac{\O}{\O_N} = 1-\frac{3k}{2}-\frac{3Ht}{2}.
\ee
The last two terms are of the same order near Earth when $t \gtrsim 100$ years.

\subsection{Holonomy of a general vector along circular orbits}\label{ss3.2}

We now turn to the case of an arbitrary vector being parallely transported along a circular orbit described by comoving radial coordinate $r = const.$ and $\theta =\pi/2$. Therefore $dr=d\theta=0$ and the above set of parallel transport equations (\ref{dAt})-(\ref{dAf}) reduce to the following:

\begin{widetext}
\bea
dA^t &+& \frac{k}{a^2r\lt(1+ \frac{k}{2a}\rt)^4}A^r dt
 + Har\lt(1+\frac{k}{2a}\rt)^2 A^\f d\f=0 \label{dAt1}\\
dA^r &+& \frac{k}{a^2r\lt(1+ \frac{k}{2a}\rt)^4}A^t dt
-\frac{\lt(1-\frac{k}{2a}\rt)}{\lt(1+\frac{k}{2a}\rt)}A^\f d\f=0 \label{dAr1}\\
dA^\th &=& 0\label{dAth1}\\
dA^\f &+& H ar \lt(1+\frac{k}{2a}\rt)^2 A^t  \, d\f + \frac{\lt(1-\frac{k}{2a}\rt)}{\lt(1+\frac{k}{2a}\rt)} A^r d\f  = 0.\label{dAf1}
\eea
\end{widetext}
From the above we can easily see that $-A^tdA^t+A^rdA^r+A^\th dA^\th+A^\f dA^\f=0$, which implies that $-(A^t)^2 + (A^r)^2 + (A^\theta)^2 + (A^\phi)^2 = constant$, as pointed out in \S\ref{sec2}. This constraint reduces the number of non-trivial independent equations to two (as we can always integrate (\ref{dAth1}) trivially).

It will be worthwhile to calculate here, how the scalar product of the arbitrary vector and the tangent vector ($X^aA_a$) changes as the former is parallely transported along a geodesic.
We have
\be
X^b\nabla_b(X^aA_a)= (X^b\nabla_bX^a)A_a+(X^b\nabla_bA_a)X^a\;.
\ee
By equations (\ref{geoeqn}) and (\ref{para}), the above equation becomes
\be\label{scalar}
X^b\nabla_b(X^aA_a)= (X^aA_a) f
\ee
Hence we see that for any arbitrary curve parameter, if these two vectors are perpendicular at a given spacetime point, they continue to be perpendicular at all points on the geodesic passing through that given point. This highlights the importance of the gyroscope, in which the spin vector is always held perpendicular to the four-velocity, as the most viable instrument to measure the holonomy of vectors.

\subsection{Holonomy in gyroscope spin}\label{ss3.3}
 As a thought experiment to directly measure holonomy, one might place a gyroscope in orbit around the central mass in a McVittie universe.  The expansion should influence the ``geodetic" precession of the gyro compared to the Schwarzschild case\cite{Hartle}.   This is not to be confused with the so-called Lense-Thirring effect, measured by LAGEOS and Gravity Probe B, which depends on the rotation rate of the central object; geodetic precession as discussed here occurs even when the central object is not rotating. As just mentioned, a gyroscope consists of a spin vector $S^a$, which is taken to be perpendicular to the tangent vector of the orbit $X^a$, such that $S_aX^a = 0$. Typically, one would find $S^a$ by solving the gyroscope equation
\be
\frac{ds^a}{d\t} + \G^a_{\ c d} S^c u^d = 0 .
\ee
However, because of the constraints present in this problem and because the spin vector is parallel transported according to the previous equations, one can employ the following procedure, which is equivalent, but somewhat simpler.

For a circular orbit the tetrad components of the four velocity were given by Eq. (\ref{tanvec}), and {because $X^a S_a = 0$} this provides a relation between $S^t$ and $S^\f$:
 \be\label{StSf}
S^t=\frac{\left(1+\frac{k}{2a}\right)^3}{\left(1-\frac{k}{2a}\right)}ar\O S^\f\;.
\ee
With Kepler's third law, Eq. (\ref{kepler}) this becomes
\be\label{StSf1}
S^t=\pm\frac{\sqrt{\frac{k}{a}}}{\lt(1-\frac{k}{2a}\rt)}S^\f
\ee
Since by definition, the vector $S^a$ is spacelike (as it is perpendicular to the 4-velocity), we can always normalise it such that $S^aS_a=1$. Due to the spherical symmetry of the problem we can also without loss of generality take $S^\theta=0$. The constraint Eq. (\ref{constraint1}) then becomes
\be\label{constraint2}
-(S^t)^2+(S^r)^2+(S^\f)^2=1\;.
\ee
Note that these two constraints, (\ref{StSf1}) and (\ref{constraint2}), reduce the number of independent non-trivial equations to one. The time evolution of $S^r$ is then given by Eq. (\ref{dAr1})  as
\be\label{Sr1}
\frac{dS^r}{dt}+\frac{k}{a^2r\lt(1+ \frac{k}{2a}\rt)^4}S^t-\frac{\lt(1-\frac{k}{2a}\rt)}{\lt(1+\frac{k}{2a}\rt)}\O S^\f=0
\ee
Inserting Eqs. (\ref{StSf1}) and (\ref{kepler}) into this expression we find after simplification
\be\label{Sr2}
\frac{dS^r}{dt}\mp\frac{\lt(1-\frac{2k}{a}+\frac{k^2}{4a^2}\rt)\sqrt{\frac{k}{a}}}{ar\lt(1-\frac{k}{2a}\rt)\lt(1+\frac{k}{2a}\rt)^4}S^\f=0
\ee
Now, substituting Eq. (\ref{StSf1}) into Eq. (\ref{constraint2}) gives
\be\label{constraint3}
S^\f=\pm\frac{\lt(1-\frac{k}{2a}\rt)}{\sqrt{1-\frac{2k}{a}+\frac{k^2}{4a^2}}}\sqrt{1-(S^r)^2}\;.
\ee
Inserting this expression into Eq. (\ref{Sr2}) yields the required decoupled equation
\be\label{Sr3}
\frac{dS^r}{dt}\mp\Psi(t)\sqrt{1-(S^r)^2}=0\, ,
\ee
where
\be\label{psi}
\Psi(t)=\frac{\sqrt{\lt(1-\frac{2k}{a}+\frac{k^2}{4a^2}\rt)\frac{k}{a}}}{ar\lt(1+\frac{k}{2a}\rt)^4}\;.
\ee
The general solution of equation (\ref{Sr3}) is given by
\be\label{Srsol}
S^r(t)=\mp\sin\left[c_1+\int_{t_0}^t \Psi(t)dt\right]\;.
\ee
Then by Eqs. (\ref{StSf1}) and (\ref{constraint3}) the other components of the spin vector are
\be\label{Sfsol}
S^\f(t)=\mp\frac{\lt(1-\frac{k}{2a}\rt)}{\sqrt{1-\frac{2k}{a}+\frac{k^2}{4a^2}}}\cos\left[c_1+\int_{t_0}^t \Psi(t)dt\right]\;,
\ee
and
\be\label{Stsol}
S^t(t)=\mp\frac{\sqrt{\frac{k}{a}}}{\sqrt{1-\frac{2k}{a}+\frac{k^2}{4a^2}}}\cos\left[c_1+\int_{t_0}^t \Psi(t)dt\right]\;.
\ee
Thus we have a complete general solution for $S^a$ in the McVittie spacetime.

In the  special case of Schwarzschild spacetime ($a=1$),
\be\label{psi0}
\Psi \equiv \Psi_0=\frac{\sqrt{k\lt(1-2k+\sfr14 k^2\rt)}}{r\lt(1+\sfr12 k\rt)^4} = const.\;.
\ee
In this case the components of the spin vector become:
\be\label{Stsol1}
S^t(t)=\mp\frac{\sqrt{k}}{\sqrt{1-2k+\sfr14k^2}}\cos\left[c_1+\Psi_0(t-t_0)\right]\;.
\ee
\be\label{Srsol1}
S^r(t)=\mp\sin\left[c_1+\Psi_0(t-t_0)\right]\;,
\ee
\be\label{Sfsol1}
S^\f(t)=\mp\frac{\lt(1-\sfr12 k\rt)}{\sqrt{1-2k+\sfr14k^2}}\cos\left[c_1+\Psi_0(t-t_0)\right]\;,
\ee

Thus, all the three spin components oscillate with constant frequency $\Psi_0$. In the limit $k<<1$, we  easily find that
\be\label{psi0lim}
\Psi_0=\O_N\lt(1-3k\rt)\;.
\ee
This expression apparently differs from the one given, for example, by Hartle \cite{Hartle} in his Eq. (14.15); however when one transforms from isotropic to Schwarzschild coordinates one finds that the two frequencies are in fact identical. In the McVittie spacetime, by contrast, we see that $a \to \infty$ implies that $\Psi \to 0$, which means that the oscillations of the spin vector are damped; $S^r$ and $S^\f$ become constant and $S^t$ goes to zero. Thus, by running placing a gyroscope in orbit around the Sun and measuring the behavior of the spin vector over cosmological times, one would certainly be able to detect the universal expansion by this method.

We now turn to two potentially more feasible experiments.

\subsubsection{Experiment with one comoving and one orbiting apparatus}\label{sec3:dual}

\begin{figure}[h]
	\centering
	\includegraphics[width=0.7\linewidth]{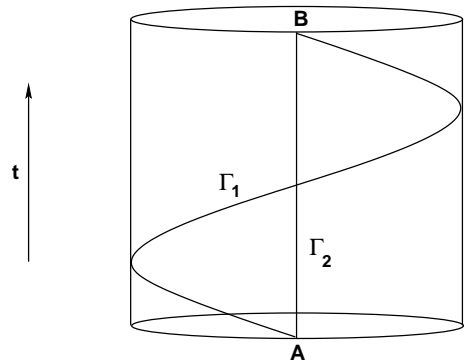}
	\caption{\textit{Spacetime diagram for the experiment}}
	\label{fig:doc3}
\end{figure}

As an alternative to the experiment just described, we can imagine one involving two gyroscopes (see Figure \ref{fig:doc3}), the first of which is transported from point A around a circular path $\G_1$ of radius $r_0$ as before, while the second follows a timelike path $\G_2$ at constant comoving radius $r_0$ and constant angular coordinates $\theta_0,\phi_0$. (Again, this is not a geodesic and rocket engine, for example, would be required to keep the apparatus in position.) These paths, coincident at time $t=t_0$,  will meet again at a point $B$ at time $t = t_0+t_{2\pi}$. The total holonomy is found by comparing the vector components at the point B where the two paths intersect again.

For the spin vector $S^a_2$, {because $d r = d\theta = d\phi = 0$,} the parallel transport equations (\ref{dAt})-(\ref{dAf}) along $\G_2$ give
\bea
dS^t_2 &+& \frac{k}{a^2r\lt(1+ \frac{k}{2a}\rt)^4}S^r_2 dt
=0\;, \label{dAt4}\\
dS^r_2 &+& \frac{k}{a^2r\lt(1+ \frac{k}{2a}\rt)^4}S^t_2 dt
=0\;, \label{dAr4}\\
dS_2^\th &=&  0\;,\label{dAth4}\\
dS_2^\f &=&   0\;.\label{dAf4}
\eea
Furthermore, since $S_aX^a=0$, and along the path $\G_2$ we have $X^a=[X^t,0,0,0]$, we must have therefore $S^t_2=0$ along the path. Plugging this into Eq.(\ref{dAr4}) shows that $S^r_2 = const.$. However, because $dS^t_2 = 0$, Eq. (\ref{dAt4}) requires $S^r_2=0$. Since without any loss of generality we can take $S^\th=0$, the normalised spin vector along the path $\G_2$ is the constant vector
\begin{equation}\label{eq:initial}
S^a_2(t)=[0,0,0,1].
\end{equation}

 Now let both the devices coincide at the same spacetime point at $t=t_0$, where we  take the readings of their corresponding spin vectors.
The first apparatus moving along $\G_1$ will obey  equations  (\ref{Srsol}, \ref{Sfsol}, \ref{Stsol}). At the initial time $t_0$, set $a(t_0)=1$. To get appropriate initial conditions we choose $c_1=\pi/2$. Then at  $t = t_0$,
\be\label{Stsol2}	 S^t_1(t_0)=\mp\frac{\sqrt{\frac{k}{a}}}{\sqrt{1-\frac{2k}{a}+\frac{k^2}{4a^2}}}\cos\left[\frac{\pi}{2}+\int_{t_0}^{t_0} \Psi(t)dt\right]=0\;.
\ee
\be\label{Srso2}
S^r_1(t_0)=\mp\sin\left[\frac{\pi}{2} +\int_{t_0}^{t_0} \Psi(t)dt\right]\;= 1,
\ee
\be\label{Sfsol2} S^\f_1(t_0)=\mp\frac{\lt(1-\frac{k}{2a}\rt)}{\sqrt{1-\frac{2k}{a}+\frac{k^2}{4a^2}}}\cos\left[\frac{\pi}{2}+\int_{t_0}^{t_0} \Psi(t)dt\right]=0\;,
	\ee

Thus at the initial time we have
	 \be
 S^a_1(t_0)=[0,1,0,0] \label{s_initial}
 \ee
and at that time the difference in the spin of these two gyroscopes is $[0,1,0,-1]$. Clearly these spins are perpendicular to each other. Ideally one would like to have these vectors be parallel to each other. However, the parallel transport equations, and the constraint that the spin vector must always be perpendicular to the 4-velocity, force parallel initial spins to be $[0,0,1,0]$.  This means that instead of taking the constant $S^\th=0$, we set the value to unity, without loss of generality. Unfortunately, this initial spin vector is a fixed point for the transport equations of both the gyroscopes. Therefore if we pick the spins to be initially parallel they remain parallel and there will be no holonomy, which is why we chose the initial conditions as above.

With the given initial conditions we next perform a measurement at the final point $B$, when the first apparatus has completed a full rotation, and the both apparatus again coincide. The time interval between these two measurement is $t_{2\p}-t_0$, where $t_{2\p}$ is the solution of
\be\label{eqnt2p}
2\pi = \frac{\sqrt{k}}{r}\int_{t_0}^{t_0+t_{2\p}} \frac{dt}{a^{3/2}\lt(1+\frac{k}{2a}\rt)^3}\,.
\ee
We can immediately see that at this point the spin vectors of the two gyroscopes are no longer perpendicular to each other.
Therefore, the net holonomy of the vector $S^a_1$ is given by
\be\label{eq:initial1}
\Delta S_1^a=S_1^a(t_0+t_{2\pi})-S_1^a(t_0).
\ee

We now consider an experiment that runs for less than cosmological times.  From Eq. (\ref{psi}), we have to order $k$
\be
\Psi(t) = \frac1{a^{3/2}}(1-\frac{3k}{a})\O_N . \label{psi1}
\ee

For $a = ((t_o + \D t)/t_o)^n$ with $\D t << t_o$, Eqs. (\ref{Srsol})-(\ref{Stsol}) then give to first order in $k$:
\begin{widetext}
\bea
S^t &=& \mp k^{1/2}(1+k - \frac{n \D t}{2 t_o})\cos \left[c_1 + (1-3k -\frac{3}{4}n \frac{\D t}{t_o})\O_N \D t\right]\\
S^r &=& \mp\sin \left[c_1 + (1-3k -\frac{3}{4}n \frac{\D t}{t_o})\O_N \D t\right]\\
S^\f &=& \mp (1+\frac{k}{2})\cos \left[c_1 + (1-3k -\frac{3}{4}n \frac{\D t}{t_o})\O_N \D t\right].
\eea
and for $a = e^{Ht}$, with $Ht << 1$ and $t_0=0$:
\bea
S^t &=& \mp k^{1/2}(1+k - \frac{H t}{2}) \cos \left[c_1 + (1-3k -\frac{3}{4}Ht)\O_N t\right]\\
S^r &=& \mp\sin \left[c_1 + (1-3k -\frac{3}{4}Ht)\O_N t\right]\\
S^\f &=& \mp (1+\frac{k}{2})\cos \left[c_1 + (1-3k -\frac{3}{4}Ht)\O_N t\right].
\eea
\end{widetext}

Formally, the time $t_{2\p}$  is given by equation (\ref{eqnt2p}). However, to the required accuracy, we may take $t_{2\p}$ to be the Newtonian value $t_{2\p} = 2\p r k^{-1/2}$, in which case $\O_N t$ in the above expressions becomes merely $2\p$.  The holonomy for the spin components is then $\D S^a = S^a (t_{2\p}) - S^a (0)$.  The obvious choices for initial conditions are $c_1=0$ and $c_1 = \p/2$.  However, the former results in initial spin components that depend on $k$, whereas the latter gives simply
$ S^a_1(t_0)=[0,1,0,0]$ as in Eq. (\ref{s_initial}).  We therefore confine ourselves to this situation, which is presumably easier to experimentally arrange.   The holonomy in the spin components for the deSitter case is then to lowest order
\bea
\D S^t &=& 6\p k^{3/2} + 3\p^2 Hr\\
\D S^r &=& 18\p^2 k^2 + 18 \p^3 k^{1/2} Hr\\
\D S^\f &=& 6 \pi k + \frac{3 \p^2 Hr}{k^{1/2}},
\eea
with similar expression for the power-law universe. Taking an orbit of 1 AU, we have $k \approx 10^{-8}$ and $Hr \approx 10^{-15}$.   The largest change in holonomy can be seen to be in the $S^\f$ component, where the first term in $\D S^\f$ is $\sim 10^{-7}$ and the second term  is $\sim 3 \times 10^{-10}$. (However in the outer reaches of the solar system, the second term dominates.) The holonomy for all the components after one rotation is
\be
\D S^a \sim \left[ 10^{-11}, 10^{-14}, 0, 10^{-7}\right].\\
\ee

More useful, however, is the deviation from the Schwarschild geometry.  The first term (independent of $H$) in each of the above expressions is the holonomy produced in Schwarzschild.  The fractional deviation of McVittie from Schwarzschild
\be
f^a = \frac{\D S^a_S - \D S^a_{Mc}}{\D S^a_S} ,\label{fractional change}
\ee
is then
\bea
f^t &=& -\frac{\p Hr}{2 k^{3/2}}  \sim 10^{-3}\\
f^r &=& -\frac{\p Hr}{ k^{3/2}} \sim 10^{-3}\\
f^\f &=& -\frac{\p Hr}{2 k^{3/2}}\sim 10^{-3}.
\eea

 Complete numerical integration agree with these results and shows that, in the scale of earth's orbit around the sun, there is virtual no difference between the power law expansion and deSitter expansion of the universe (see Figure 2). The total holonomy over one complete rotation found numerically is shown in the Table in \S\ref{discussion}.

Although the holonomy produced by this experiment is in principle detectable, with advanced enough technology, the problem of keeping the `stationary' observer at the same comoving radius probably makes this proposal unviable even in principle. Is there a simpler proposal? We turn to one possibility now.

\subsubsection{Experiment with two counter orbiting gyroscopes}\label{sec:3_counter}

We reiterate that to compute holonomy one must compare the components of the tangent vector after the particle or measuring apparatus has traversed a closed path.  Even in Schwarzschild the comparison cannot be made with a single apparatus because after one orbit the instrument is no longer in its original spacetime position. We can, however, imagine two devices with their spin vectors aligned at time $t = 0$, which are then sent out on circular orbits as above in opposite directions through an angle $2\pi$, to meet at the same comoving observation point. Overall, this combination of curves gives a closed orbit that can be used to measure holonomy. Since the two devices have an $\O$ of the same magnitude but opposite sign at every point, the changes in the vector components will add, and so the total holonomy of each will be half of their sum.

Geometrically, the picture of this holonomy is as follows: At the initial time, $S^a(t_0)=\left[S^t(t_0),S^r(t_0),0,S^\f(t_0)\right]$, while at $t=t_{2\p}$, the vector becomes $S^a(t_{2\p})=\left[S^t(t_{2\p}),S^r(t_{2\p}),0,S^\f(t_{2\p})\right]$. This represents boosts in two different directions relative to the tetrad frame, which by (\ref{eq:const}), because the magnitude of the vector remains unchanged, is a Lorentz transformation. The sum of the boosts in different directions represents a rotation. Since $S^\th = 0$, the net change of the spatial part of the vector $S^a$ lies in the $(r,\f)$ plane, in other words, the $\theta$ direction.

 This can be seen more formally by examining the Lorentz Group commutation relations
\be
[J_i,J_j]=i\e_{ijk}J_k,
\ee
\be
[J_i,M_j]=i\e_{ijk}M_k,
\ee
\be
[M_i,M_j]=-i\e_{ijk}J_k  .
\ee
Here, $J_i$ is the generator of rotations $M_i$ is the generator of boosts. The last commutator shows the spatial holonomy will be in the $\theta$ direction, as stated.\\

The change in the vectors are the same as in the experiment in the previous section, but in this case the the net holonomy change will be  $2\Delta S^{a}$. Hence, as before, on the scale of the solar system there will be no difference between the power law expansion and deSitter expansion of the universe, and the order of the net holonomy remains $\D S^a \sim \left[ 10^{-11}, 10^{-14}, 0, 10^{-7}\right]$.

\begin{figure}[!tb]
	\centering
\vbox{\hfil\scalebox{.45}
{\includegraphics{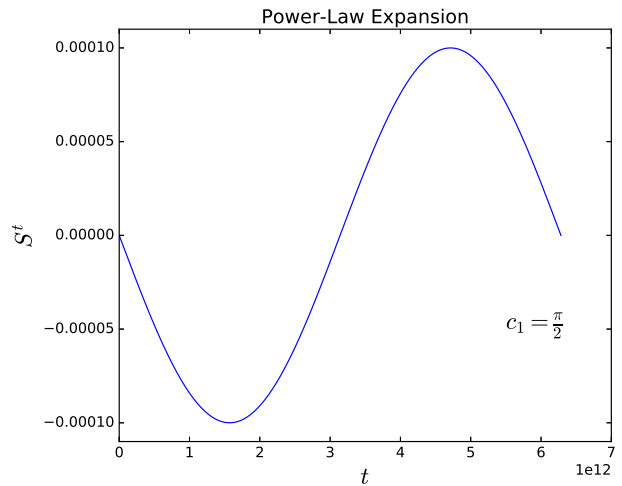}}\hfil}
	\caption{\textit{Spin vector solution $S^{t}$ for power law expansion, $a(t)=(t/t_{0})^{2/3}$. Numerically, the net holonomy in this component is calculated to be of the order $10^{-11}$}.}
	\label{fig:Stpowerlaw}
\end{figure}

\begin{figure}[htb]
	\centering
	\includegraphics[width=8.5cm, height=7cm]{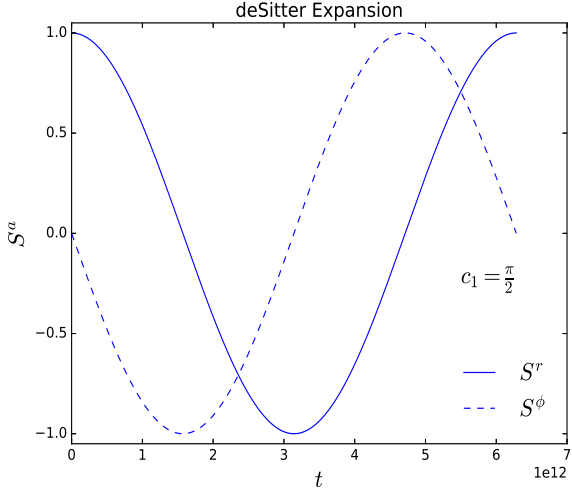}
	\caption{\textit{Spin vector solutions for deSitter expansion, $a(t)=e^{Ht}$, where $H=const$. These solutions are identical to the power law expansion, using the same initial conditions as in that case}.}
	\label{fig:SdeSitter}
\end{figure}

\begin{figure}[htb]
	\centering
	\includegraphics[width=8.5cm, height=7cm]{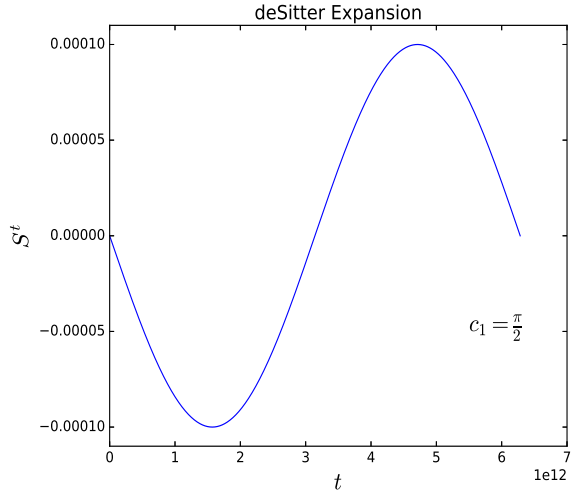}
	\caption{\textit{$S^{t}$ solution for deSitter expansion. This solution is also identical to the power law expansion case}}
	\label{fig:Stpowerlaw}
\end{figure}

\begin{figure}[htb]
	\centering
	\includegraphics[width=8.5cm, height=7cm]{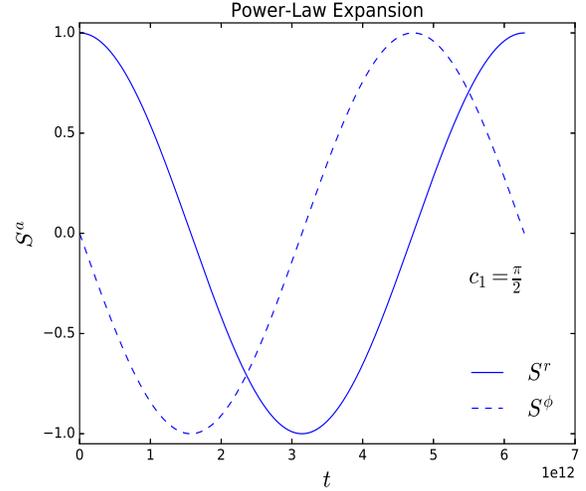}
	\caption{\textit{Spin vector solutions, $S^{r}$ and $S^{\phi}$, with $k=10^{-8}$ and $Hr=10^{-15}$ as the initial conditions for the power law expansion. Eq. (\ref{psi}) is used as the expression for $\Psi(t)$, and we find the net holonomy in the spin components to be the same as predicted when taking the expression for $\Psi(t)$ to first order in $k$, Eq. (\ref{psi1}).}}
	\label{fig:powerlaw}
\end{figure}

\subsubsection{Error introduced by assuming circular geodesics}\label{sec:3_error}

We emphasized at the beginning of \S\ref{ss3.1} that circular orbits in the McVittie spacetime are not actually geodesics.  During the course of an orbit, freely falling instruments spiral inward from $r_1$ to $r_2$.  If we do not wish to use rockets to hold the apparatus at a constant $r$, we therefore cannot compute the holonomy after a complete orbit.  However, it is easy to see that the error introduced by assuming the orbit is circular, as we have done, is negligible.

With $k \equiv m_o/r$ we found, for example, in the $c_1 = \pi/2$ de Sitter case that
\be
\D S^r = 18\p^2 k^2 + 18 \p^3 k^{1/2} Hr \label{DSr1}
\ee
for circular orbits.

Now, instead, take two circular orbits at $r_2$ and $r_1$. If as above
\be
f_1 = \frac{\D S^t(r_1)_S - \D S^t(r_1)_{Mc}}{\D S^t(r_1)_S}
\ee
\be
f_2 = \frac{\D S^t(r_2)_S - \D S^t(r_2)_{Mc}}{\D S^t(r_2)_S},
\ee
then the quantity of interest to measure the fractional deviation of McVittie from Schwarzschild by using $r_2$ instead of $r_1$ is evidently
\be
\D f = \frac{f_2 -f_1}{f_1}.  \label{df}
\ee

In our notation, McVittie shows that in the lowest approximation
\be
\frac1{r} = \frac{m_o a(t)}{h^2} ,
\ee
where $h$ is some constant.  If $a = 1$ at $t = 0$,
\be
\frac1{r_1} = \frac{m_o}{h^2} .
\ee
Taking $\D r \equiv r_2 - r_1$ we have for  $a = e^{Ht}$
\be
-\frac{\D r}{r_1} = H \D t << 1.
\ee
(Indeed, with the parameters used previously, $\D r$ is of the order 10 meters.) Eqs. (\ref{DSr1}) and (\ref{df}) then gives
\be
\D f^r = 2 H\D t = \frac{4\p H r_1}{k^{1/2}}.
\ee
As this number $\sim 10^{-10}$, we see that the error introduced in computing the holonomy by ignoring the difference between $r_2$ and $r_1$ is completely negligible.

{The results of this section suggest that a simpler experiment might be merely to put a satellite in orbit around the Sun and measure $\D r$, avoiding a spin measurement. In that sense such an experiment is indeed simpler.  However, one would need to hold some device (a second satellite) at the original $r$ in order to make the distance measurement, which would require a force.  Secondly, for an orbit at 1 AU, $\D r/(orbital \ circumference) \sim 10^{-11}$, suggesting that the precision required for such a measurement is no less than for the gyro experiment.}

\section{Discussion}\label{discussion}
We have discussed a few simple thought experiments, which at first sight might in principle actually be performed. These have been carried out in the McVittie spacetime, which assumes that the ``solar system" (the Sun and a test particle) is directly embedded in an expanding universe. Employing the McVittie metric has enabled us to calculate the holonomy produced in gyroscopes on solar-system scales. One might perform similar calculations for other spacetimes as well, for example the Tolman models. However none of those models contain the limiting cases of pure Schwarzschild geometry on the one side and pure FLRW geometry on the other, which the McVittie spacetime has. Hence the results in those cases would be more unrealistic.

In terms of holonomy, the key difference between a pure Schwarzschild geometry and the McVittie spacetime that emerges from our calculations is the variation in both amplitude and frequency in the oscillations of a gyroscope's spin vectors. The table below summarizes the numerical results of the experiments we have discussed, including the fractional change of components of the spin vector between the two spacetimes.

\begin{center}
\begin{table}
\begin{tabular}{||c|c|c|c||}
	\hline
	\hline
 	& Schwarzschild & McVittie & Fractional change\\
	\hline
$\D S^t$ & $ 1.885\times 10^{-11}$ & $1.888\times 10^{-11}$ & $-1.592 \times 10{-3}$  \\
	\hline
$\D S^r$ & $-1.776\times 10^{-14}$ & $-1.787\times 10^{-14}$ & $-6.194 \times 10^{-3}$\\
	\hline
$\D S^\f$ & $1.885\times 10^{-7}$ & $-1.188\times 10^{-7}$ & $-1.592\times 10^{-3}$\\
\hline
\hline
\end{tabular}
\caption{The fractional deviation of the Schwarzschild geometry from the McVittie spacetime, as defined by Eq.(\ref{fractional change}), for the deSitter case when $c_{1}=\pi/2$.}
\end{table}
\end{center}

{Of course, the real universe does not behave like the McVittie spacetime.  The solar system contains nine planets.  However, by far most of the mass is concentrated in Jupiter, which is only $\sim 10^{-4}$ M$_{\odot}$.  As Jupiter is also much farther away from Earth than the Sun, $k_{J} \sim 10^{-5} k_{\odot}$ and so any perturbation to the metric would be extremely small.  In any case, since we are interested in the difference in measurements between the Schwarzschild and McVittie spacetime, any such perturbation would at least to first order subtract out.}

{The main problem concerns  the  scale at which static or quasi-static domains coalesce out of the expanding universe as structure formation takes place. This is essentially the issue of virialization of emerging gravitational structures\cite{peebles}. Given that virialization takes place on the scale of galactic clusters, this is in principle the scale on which one would have to carry out realistic experiments. As the order of magnitude of that holonomy we have calculated is only $\sim 10^{-11}$, to distinguish the two geometries does not appear feasible on solar system scales because the virialization scale is so much larger.  Nevertheless, in the tradition of Einstein and Strauss\cite{ES45} and Noerdlinger and Petrosian\cite{NP71}, it is conceptually interesting to consider idealized experiments that show effects on solar-system scales,
and even more so in the context of any consideration of how the universe at large influences local physics.}

\section{Acknowledgements}
RG, GE and MC thank the South African National Research Foundation (NRF) for support, and GE thanks the University of Cape Town Research Committee (URC) for financial  support.

\end{document}